\documentclass{aa}
\usepackage{graphicx}
\usepackage{txfonts}
\usepackage{natbib}
\usepackage{siunitx}

\begin{document}
\title{Observations and analysis of phase scintillation of spacecraft signal on the interplanetary plasma}

\author{G. Molera~Calv\'es\inst{1,2}
  \and S.~V. Pogrebenko\inst{1}
  \and G. Cim\`o\inst{1}
  \and D.~A. Duev\inst{1,3}
  \and T.~M. Bocanegra-Baham\'on\inst{1,4,5}
  \and J.~F. Wagner\inst{6,2}
  \and J. Kallunki\inst{2}
  \and P. de~Vicente\inst{7}
  \and G. Kronschnabl\inst{8}
  \and R. Haas\inst{9}
  \and J. Quick\inst{10}
  \and G. Maccaferri\inst{11}
  \and G. Colucci\inst{12}
  \and W.~H. Wang\inst{5}
  \and W.~J. Yang\inst{13}
  \and L.~F. Hao\inst{14}
  }

\institute{Joint Institute for VLBI in Europe, Oude Hogeevensedijk 4, 7991PD, Dwingeloo, The Netherlands.
 \email{molera@jive.nl}
 \and Aalto University Mets\"ahovi Radio Observatory, Kylm\"al\"a, Finland.
 \and Sternberg Astronomical Institute, Lomonosov Moscow State University, Russia.
 \and Department of Astrodynamics and Space Missions, Delft University of Technology, The Netherlands.
 \and Shanghai Astronomical Observatory, Chinese Academy of Sciences, Shanghai 200030, China.
 \and Max Planck Institute for Radio Astronomy, Bonn, Germany.
 \and National Astronomical Observatory, Astronomical Centre of Yebes, Spain.
 \and Federal Agency for Cartography and Geodesy, Geodetic Observatory of Wettzell, Germany.
 \and Chalmers University of Technology, Onsala Space Observatory, Sweden. 
 \and Hartebeesthoek Radio Astronomy Observatory, Krugersdorp, South Africa.
 \and National Institute for Astrophysics, RadioAstronomy Institute, Radio Observatory Medicina, Italy. 
 \and E-geos S.p.A, Space Geodesy Center, Italian Space Agency, Matera, Italy.
 \and Xinjiang Astronomical Observatory, Chinese Academy of Sciences, Urumqi 830011, China.
 \and Yunnan Astronomical Observatory, Chinese Academy of Sciences, Kunming 650011, China.
}

\date{Received 28 October 2013 / Accepted 05 February 2014}

  \abstract
   {}
   {The phase scintillation of the European Space Agency's (ESA) Venus Express (VEX) spacecraft telemetry signal was observed at X-band ($\lambda=3.6\,$cm) with a number of radio telescopes of the European VLBI Network (EVN) in the period $2009$--$2013$.}
   {We found a phase fluctuation spectrum along the Venus orbit with a nearly constant spectral index of $-2.42\,\pm\,0.25$ over the full range of solar elongation angles from $0^\circ$ to $45^\circ$, which is consistent with Kolmogorov turbulence. Radio astronomical observations of spacecraft signals within the solar system give a unique opportunity to study the temporal behaviour of the signal's phase fluctuations caused by its propagation through the interplanetary plasma and the Earth's ionosphere. This gives complementary data to the classical interplanetary scintillation (IPS) study based on observations of the flux variability of distant natural radio sources.}
   {We present here our technique and the results on IPS. We compare these with the total electron content (TEC) for the line of sight through the solar wind. Finally, we evaluate the applicability of the presented technique to phase-referencing Very Long Baseline Interferometry (VLBI) and Doppler observations of currently operational and prospective space missions.}
   {}
\maketitle 

\section{Introduction}
The determination of the Doppler parameters and state vectors of spacecraft by means of radio interferometric techniques opens up a new approach to a broad range of physical processes. The combination of Very Long Baseline Interferometry (VLBI) and Doppler spacecraft tracking has been successfully exploited in a number of space science missions, including tracking of VEGA balloons for determining the wind field in the atmosphere of Venus~\citep{Preston}, VLBI tracking of the descent and landing of the Huygens Probe in the atmosphere of Titan~\citep{Bird}, VLBI tracking of the impact of the European Space Agency's (ESA) Smart-$1$ Probe on the surface of the Moon with the European VLBI Network (EVN) radio telescopes, and the recent VLBI observations of ESA's Venus Express (VEX)~\citep{Duev}, and of the Mars Express (MEX) Phobos flyby~\citep{Molera}.

The Planetary Radio Interferometry and Doppler Experiment (PRIDE) is an international enterprise led by the Joint Institute for VLBI in Europe (JIVE). PRIDE focusses primarily on tracking planetary and space science missions through radio interferometric and Doppler measurements~\citep{Duev}. PRIDE provides ultra-precise estimates of the spacecraft state vectors based on Doppler and VLBI phase-referencing~\citep{Beasley} techniques. These can be applied to a wide range of research fields including precise celestial mechanics of planetary systems, study of the tidal deformation of planetary satellites, study of geodynamics and structure of planet interiors, characterisation of shape and strength of gravitational field of the celestial bodies, and measurements of plasma media properties in certain satellites and of the interplanetary plasma. PRIDE has been included as a part of the scientific suite on a number of current and future science missions, such as Russian Federal Space Agency's RadioAstron, ESA’s Gaia, and Jupiter Icy Satellites Explorer (JUICE).

The study of interplanetary scintillation (IPS) presented in this paper was carried out within the scope of the PRIDE initiative via observations of the ESA’s VEX spacecraft radio signal. Venus Express was launched in 2005 to conduct long-term in-situ observations to improve understanding of the atmospheric dynamics of Venus~\citep{Titov}. The satellite is equipped with a transmitter capable of operating in the S and X-bands ($2.3\,$GHz, $\lambda=13\,$cm, and $8.4\,$GHz, $\lambda=3.6\,$cm, respectively). Our measurements focussed on observing the signal transmitted in the X-band. The VEX two-way data communication link has enough phase stability to meet all the requirements for being used as a test bench for developing the spacecraft-tracking software and allowing precise measurements of the signal frequency and its phase. During the two-way link, the spacecraft transmitter is phase locked to the ground-based station signal, which has a stability, measured by the Allan variance, better than $10^{-14}$ in $100$--$1000$ seconds. The added Allan variance of the spacecraft transmitter is better than $10^{-15}$ in the same time span. The signal is modulated with the data stream. The modulation scheme leaves $25\%$ of the power in the carrier line, which is sufficient for its coherent detection. We used observations of the VEX downlink signal as a tool for studying IPS.

In this paper, we describe the solar wind and the nature of the interplanetary scintillation by analysing the phase of the signal transmitted by a spacecraft in Sect.~\ref{sec:the}. The observational setup at the radio telescopes, a short description of the tracking software, and analysis of the phase fluctuations are summarised in Sect.~\ref{sec:met}. Results from analysis of the phase scintillation in the signal from VEX are presented in Sect.~\ref{sec:res}. Finally, conclusions are presented in Sect.~\ref{sec:con}.

\section{Theory and observable values}\label{sec:the}

The solar wind is composed of an almost equal amount of protons and electrons with a lower number of heavier ions. The solar wind expands radially outwards, and each plasma particle has its own magnetic field. During the expansion of the outflow in the heliosphere, strong turbulence, which affect the propagation of the solar wind, are generated. Such turbulence resembles the well-known hydrodynamic turbulence described by~\citet{Kolmogorov}. The acceleration of the solar wind is attributed to the heating of the corona shell of the Sun. However, the physical processes of this phenomenon are not understood very well.

One method of classifying the solar wind is based on its speed. Fast winds are seen above coronal holes and can last for several weeks. They have an average velocity of~$750\,$km\,s$^{-1}$ with a low plasma particle density of~$3$\,cm$^{-3}$. With an average speed between~$400$ and~$600\,$km\,s$^{-1}$, winds are usually classified as intermediate and associated with stream interaction regions or coronal mass ejections (CME). Finally, slow winds are related to helmet streamers near the current sheet of the Sun. They have an average speed between $250$ to $400\,$km\,s$^{-1}$ and a higher plasma particle density of~$10$\,cm$^{-3}$. The speed and density of the solar wind has been constantly monitored by a number of satellites at various distances from the Sun~\citep{Phillips,Schwenn}.

Variations in the flux density of a radio wave propagating in the solar system can be associated with IPS. The radio waves emitted by a source scintillate owing to the electron density variations in an ionized plasma, such as the solar wind. A number of studies have been conducted to prove and quantify the effect of the solar wind on radio signals~\citep{Hewish,Coles,Coles2,Canals}. In these studies, the flux intensity fluctuations of the radio signal emitted by known celestial sources have been thoroughly studied at frequency bands below $2\,$GHz. Currently, several groups are measuring interplanetary scintillation. For instance, IPS and CME measurements with the Ooty radio telescope in India~\citep{Manoharan}, systematic observations of celestial sources at $\sim930$ and $1420\,$MHz with the UHF antennas of the European Incoherent SCATter radar (EISCAT,~\citealt{Fallows0,Canals,Fallows}), and 3D reconstruction of the solar wind structure with data from the Solar-Terrestrial Environment Laboratory (STELab,~\citealt{Bisi2009,Jackson}).

The work presented here approaches the scintillation in a different way from these studies. Instead of using the intensity of the signal, we derived information on the interplanetary plasma from phase fluctuations of the spacecraft signal. Unlike the case of intensity variations, measurements of phase scintillation provide information on the full range of scale sizes for the electron density fluctuations at any distance from the Sun. This technique is only feasible with precise information on the phase of the radio signal. Two methods can guarantee the necessary accuracy using spacecraft. The one-way mode, where the observations rely on an ultra stable oscillator on-board the spacecraft, and the two-way mode, where an initial modulated signal is transmitted from the ground-reference antenna, locked in the spacecraft and returned back to the Earth~\citep{Asmar}.

Several examples in the literature compare the interplanetary plasma irregularities with those present in other turbulent ionized gases~\citep{Yakovlev,Bruno,Little}. The plasma medium is considered randomly inhomogeneous within a wide range of scales. The characteristic size of inhomogeneity of the plasma can range from tens to thousands of kilometres. The relative motion of these irregularities with respect to the observer is indeed the origin of the fluctuations in the detected phase. The level of scintillation is categorised as strong or weak depending on the level of variations in the detected signal. The impact of the solar wind on the amplitude and phase of the radio waves is usually expressed in terms of a scintillation model. This can be modelled as the sum of the effect of a series of thin screens between the source and the observer. This model, known as the Born assumption~\citep{Born}, is only applicable in weak regimes, which is the case of the results presented in this paper.

The variance of the fluctuations~($\sigma_{\gamma}$) in the phase of the radio wave~($\phi$) can be calculated as the integral along the path spacecraft-to-Earth and the wavenumber of the plasma inhomogeneity spectrum~($\kappa$):

\begin{equation}
   \sigma_{\gamma}^2= (2 \pi k)^2 \int_{0}^{L} \int_{0}^{\kappa_{0}} \phi(\kappa)\kappa d\kappa  dr,
   \label{eq:sigma_gamma}
\end{equation}

\noindent where $\kappa$ ranges from $0$ to $\kappa_{0}=2\pi\Lambda_{0}^{-1}$, $k$ is the free space wavenumber, $L$ the distance spacecraft-to-Earth, and $\Lambda_{0}$ the outer turbulence scale~\citep{Yakovlev}.

In our case, to estimate the variance of the phase fluctuations, we calculate the spectral power density. By visual inspection of the spectrum (see Fig.~\ref{fig:spectra}), we define the scintillation and noise bands. Thus, the phase scintillation index~($\sigma_\mathrm{Sc}$) is estimated as the standard deviation of the phase fluctuations caused by the interplanetary plasma~($\sigma_\mathrm{Sc}=stdev(\phi_\mathrm{Sc})$), where~$\phi_\mathrm{Sc}$ is the phase fluctuations of the signal within the scintillation band. The level of the system noise fluctuations in the phase of the signal is determined as the standard deviation of the phase fluctuations within the system noise band. Both scintillation and noise bands depend mainly on the system noise of the receiver and the level of variations in the signal. The phase scintillation is given by

\begin{equation}
 \phi_\mathrm{Sc} = \left[ \int_{f_{0}}^{f_\mathrm{max}} D(f_{0}) \cdot \left( \frac{f}{f_{0}} \right)^m df \right]^{1/2},
 \label{eq:fluctuations}
\end{equation}

\noindent where $D(f_{0})$ is the spectral power density, $m$ the slope of the spectral power density, $f$ the frequency of the fluctuations, and the lower limit of integration is defined as $f_{0}=1/\tau$ with $\tau$ being the length of the phase-referencing nodding cycle~\citep{Duev}.

The total electron content (TEC) is an estimation of the amount of electrons along the line of sight (from spacecraft to the Earth-based station). To model the TEC, we do not take into account that the signal propagates along different paths in the medium (so-called Fresnel channels). At any point in the solar system, the electron density of the solar wind depends on the distance with respect to the Sun:

\begin{equation}
 Ne(r) = n_{0} \cdot \left(\frac{r_{0}}{r}\right)^{2},
 \label{eq:elec_density}
\end{equation}

\noindent where $n_{0}$ is the nominal electron density at a distance $r_{0}$. The density will depend on the propagation speed of the solar wind. In regimes closer to the Sun, higher power terms may be included in Eq.~\ref{eq:elec_density}~\citep{Yakovlev}. However, for the purpose of this study they are not included here. In our approximation, the TEC along the line of sight is therefore estimated as

\begin{equation}
 TEC = 2~tecu^{-1} \int_\mathrm{Earth}^\mathrm{S/C} Ne\left(r(l)\right) dl,
 \label{eq:TEC}
\end{equation}

\noindent where $tecu$ (TEC unit) is the total electron content per square metre and is equal to $10^{16}\,$m$^{-2}$. A factor of $2$ is used because of the two-way link. The radio signal is affected by the medium during the uplink (Earth-to-spacecraft) and the downlink (spacecraft-to-Earth-based station). 

\begin{figure}[!ht]
 \begin{center}
  \includegraphics[width=250pt]{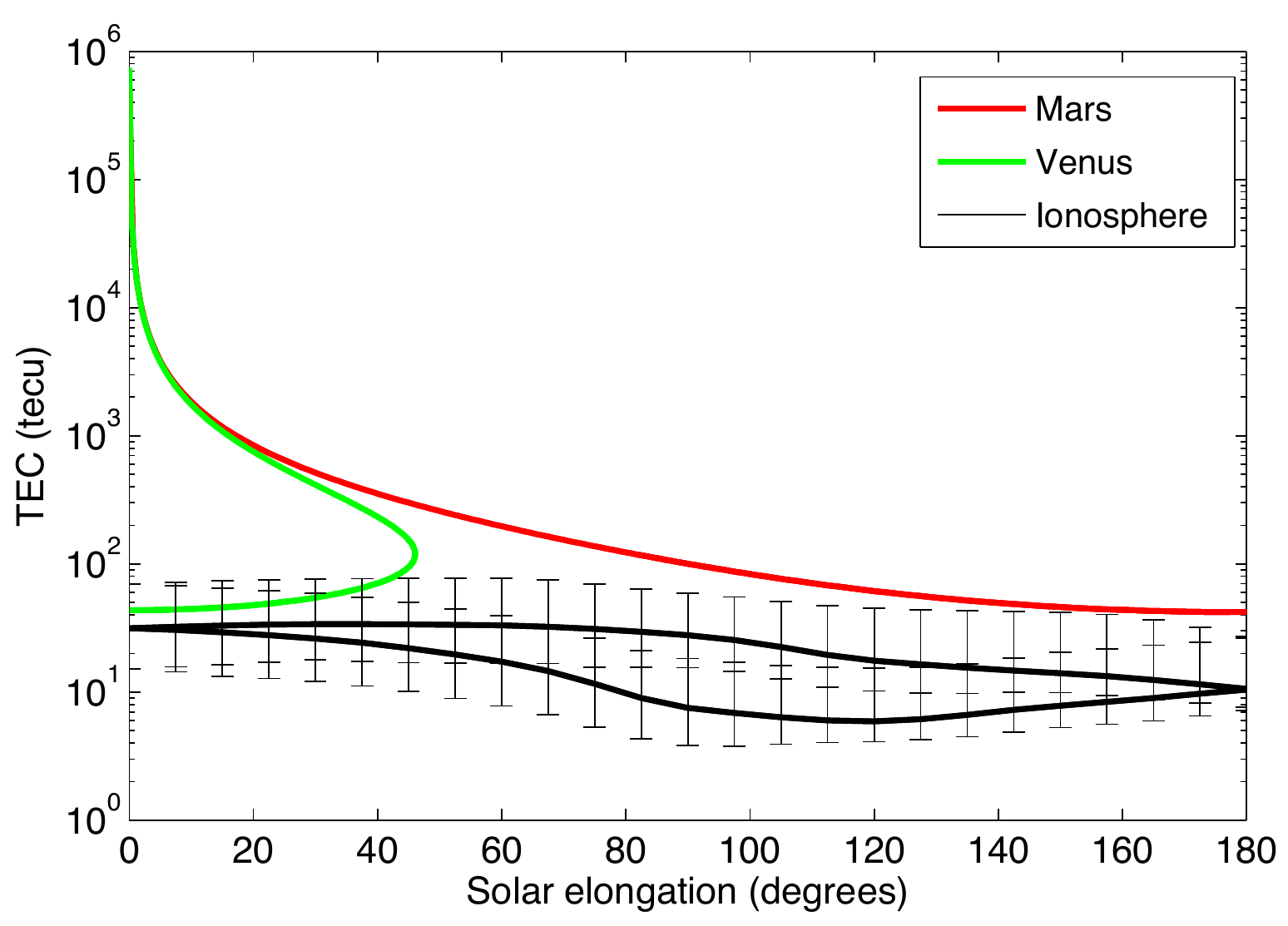}
 \end{center}
 \caption{Simulated TEC as a function of the solar elongation for a spacecraft transmitting one-way link from Venus (green), Mars (red), and Earth's ionosphere towards Venus (black). The mean value and the error bars show the maximum and minimum values of the ionosphere depending on the Earth’s orientation and Venus orbit.}
 \label{fig:tec}
\end{figure}

We calculated the TEC for the heliocentric orbits of the Earth, Venus, and Mars for the specific cases of the MEX and the VEX spacecraft. We used an electron density ($n_{0}$) of five electrons per cm$^{-3}$, which is defined as nominal at $1\,$AU in~\citet{Axford}. Our comparison between the TEC along the line of sight and the phase scintillation indices (presented in the Results) takes the electron density in fast and slow wind into account. The calculated TEC values for the Mars and Venus orbits are shown with respect to solar elongation angle in Fig.~\ref{fig:tec}. Although this work focusses on VEX, we included data for Mars since MEX can also be used for interplanetary plasma studies in the same way. The ionosphere's TEC (black line) was calculated for the last three years of observations as suggested in~\citet{Duev}.

\section{Observational summary}\label{sec:met}
\subsection{Spacecraft observations}
We conducted more than one hundred observations of the VEX spacecraft with a number of radio telescopes. The first observing session was carried out in November $2009$. The initial goal was to better understand the nature of the phase fluctuations simultaneously detected at multiple radio telescopes. This research aimed at analysing the behaviour of the fluctuations of the radio waves with respect to solar elongation. The campaign has been extended until $2014$, covering two revolutions of Venus around the Sun. Venus crossed superior conjunction in January $2010$, in August $2011$, and the last one in March $2013$.

The hardware and the observational setup of the antennas are similar for most of the VLBI radio telescopes. As a result, the observations can be conducted with any of the antennae. The main specifications of the EVN and Matera radio telescopes, used in our observations, are shown in Table~\ref{tab:antenna}. The data are recorded at each telescope in standard VLBI format. Precise time stamps allow the telescope data to be correlated. The spacecraft-tracking software used in this work was developed to be compatible with the VLBI format.

\begin{table}[!ht]
\caption{Specifications of the radio telescopes: Antenna, country code, latitude, longitude, elevation [m], and antenna diameter (\o) [m].}
\begin{center}
\small
\begin{tabular}{l|c|c|c|c|c}
Antenna			&  Code	&		Latitude			&		Longitude		& Elev. &	\o \\
\hline \hline
Hartebeeshoek	&ZA&-36$^{\circ}$26’05”	&	27$^{\circ}$41’05”	&	1415	&	26 \\
Kunming			&CN&	25$^{\circ}$01’38”	&  102$^{\circ}$47’45”	&	1973	&	40 \\	
Matera			&IT&	40$^{\circ}$38’58”	&	16$^{\circ}$42’14”	&	543	&	20 \\
Medicina			&IT&	44$^{\circ}$31’14”	&	11$^{\circ}$38’49”	&	67	&	32 \\
Mets\"ahovi		&FI&	60$^{\circ}$13’04”	&	24$^{\circ}$23’25"	&	75	&	14 \\
Noto				&IT&	36$^{\circ}$42’34”	&	14$^{\circ}$59’20”	&	143	&	32 \\
Onsala			&SE&	57$^{\circ}$23’47”	&	11$^{\circ}$55’39”	&	10	&	20 \\
Sheshan			&CN&	31$^{\circ}$05’57”	&  121$^{\circ}$11’58”	&	29	&	25 \\
Urumqi			&CN&	43$^{\circ}$28’17”	&	87$^{\circ}$10’41”	&	2033	&	25 \\
Wettzell			&DE&	49$^{\circ}$08’42”	&	12$^{\circ}$52’03”	&	670	&	20 \\
Yebes			&ES&	40$^{\circ}$31’27”	&	-3$^{\circ}$05’22”	&	999	&	40 \\
\end{tabular}
\end{center}
\label{tab:antenna}
\end{table}

More than $50\%$ of the sessions in this campaign were conducted with the Mets\"ahovi radio telescope located in Finland. The antenna has a $14$-metre dish with a System Equivalent Flux Density (SEFD) at an X-band of $3000\,$Jy. The remaining observations were carried out with the eleven other radio telescopes. Their sensitivities are slightly better than the one offered by the Finnish radio telescope (see Table~\ref{tab:antenna}). The observations were conducted in single-dish and VLBI phase-referencing modes.

In the VLBI phase-referencing technique, a strong compact radio continuum source (e.g. a quasar) with a well-determined position is used to calibrate against fast atmospheric changes and allow the coherent detection of a target source (e.g. the spacecraft). In addition, it provides an absolute reference point for determining the position of the spacecraft. The telescopes nod between the spacecraft and the reference source. For an optimal correction using phase-referencing, the sources should be close by (a few degrees apart on the plane of the sky). The nodding cycle should ensure the detection of both sources, and it is constrained by the distance between sources and by the scintillation level introduced by the interplanetary medium. The latter can be quantified by studying the propagation of the spacecraft signal in the solar wind, as described in this paper.

Our observing sessions lasted usually about three to five hours. The duration was constrained by the visibility of the target, the transmitting slots of the spacecraft, and the availability of the telescopes. The recording time ($\Delta\,T$) should consider that $\Delta\,T\,>\,\Lambda_{0}\,V^{-1}$ where $V$ is the velocity of the solar wind ($400\,$km\,s$^{-1}$) and $\Lambda_{0}$ the outer turbulent scale ($3\cdot 10^{6}\,$km), then $\Delta\,T$ should be longer than two hours~\citep{Yakovlev}. However, to simplify recording and re-pointing of the antenna, the observations were split into $19$-minute scans. These were sufficient to analyse the behaviour of the phase fluctuations with a milli-Hz resolution. The scans were then statistically averaged.

The down-converted radio signal was recorded using standard VLBI data acquisition systems: Mark5 A/B, developed at MIT/Haystack~\citep{Whitney}, or the PC-EVN, developed at the Mets\"ahovi radio observatory~\citep{Mujunen}. Both systems recorded the data onto disks using the intermediate frequency channels. We used four or eight consecutive frequency channels, with $8$ or $16\,$MHz bandwidth and two-bit encoding for a total data aggregate output ranging from $128\,$Mbps to $512\,$Mbps. The large volume of the data was constrained by the VLBI hardware. The astronomical data were stored in large raw-data files, which contained up to $34\,$GB. These files were then transferred electronically to JIVE for processing and post-analysis. Owing to the high-speed fibre connections at the radio telescopes, the observed data could be processed within an hour after the completing the session.

\subsection{Spacecraft narrowband signal processing}\label{sec:sc}
\begin{figure}[!ht]
 \begin{center}
 \includegraphics[width=250pt,clip]{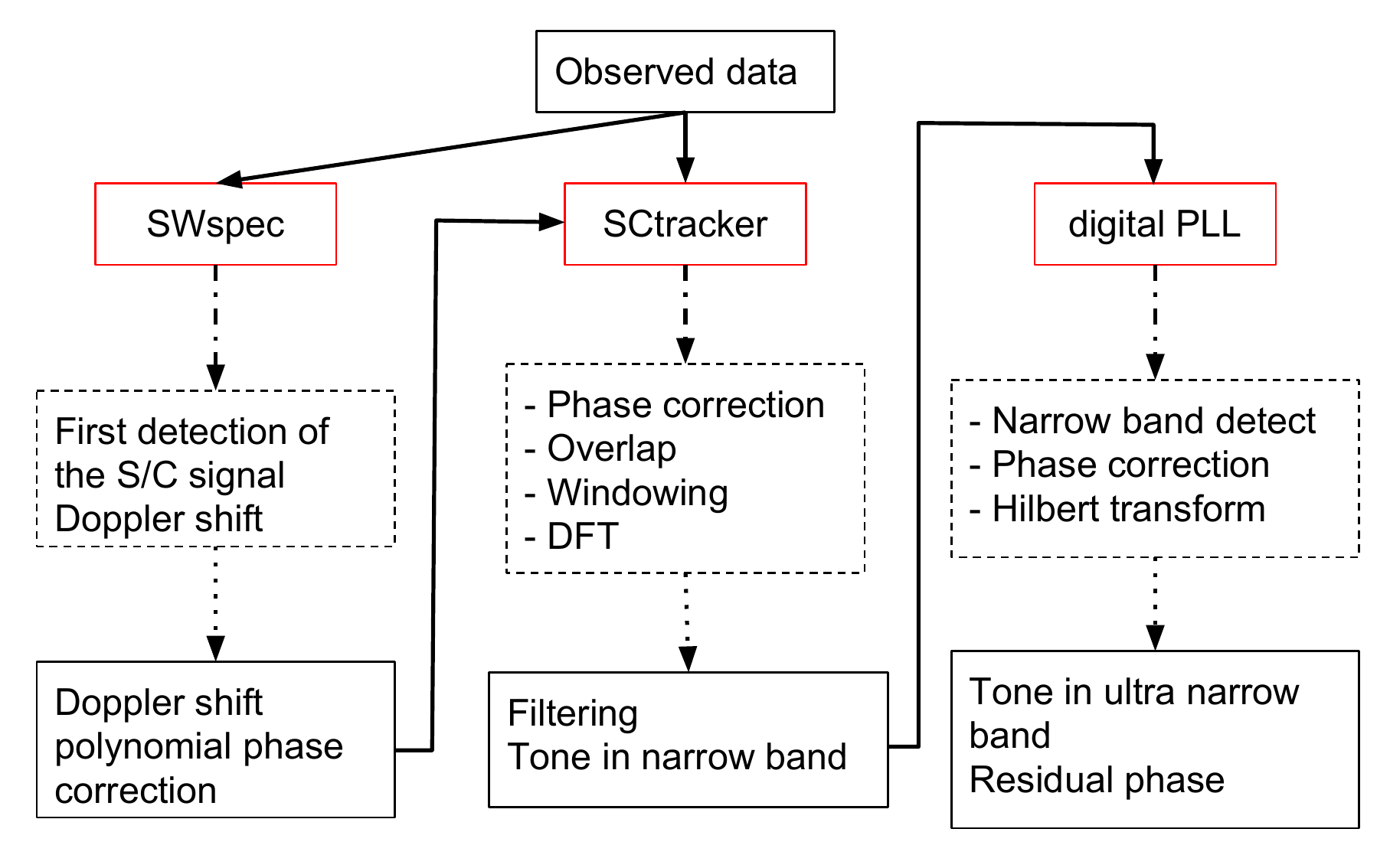}
 \end{center}
 \caption{This block diagram summarises the pipeline of the narrowband signal. It includes a brief summary of the three developed software packages: SWspec, SCtracker, and digital PLL.}
 \label{fig:pipeline}
 \end{figure}

The data-processing pipeline of the spacecraft's narrowband signal analysis consists of three different software packages: SWspec, SCtracker, and digital-PLL~\citep{Molera}. These were developed exclusively for spacecraft-tracking purposes. The processing pipeline and a brief description of the purpose of the packages are shown in Fig.~\ref{fig:pipeline}.

The high-resolution spectrometer software (SWspec) performs an accurate Windowed-OverLapped Add (WOLA) Discrete Fourier Transform (DFT) and the time integration of the spectrum. It estimates the initial detection of the spacecraft carrier signal and the adjacent tones. These detections allow us to derive the Doppler variation of the spacecraft signal along the entire scan. The Doppler shift is modelled using an $n$-order frequency polynomial fit, which is equivalent to an ($n+1$)-order phase-stopping polynomial fit. For scintillation analysis we use an $n=5$ order fit. The order of fit will also limit the frequency cut-off ($f_{\mathrm{c}}$) of the spectral power density of the phase fluctuations. 
 
The core of the narrowband processing software is the spacecraft tone tracking software (SCtracker). It applies a double-precision polynomial evaluation to the baseband sample sequence to stop the carrier tone phase and to correct for the Doppler shift. The result is the narrower band (bandwidth of $2$--$4\,$kHz) of the original signal filtered out into continuous complex time-domain signals using a second-order WOLA DFT-based algorithm of the Hilbert transform approximation.

Finally, the phase-lock-loop (PLL) calculates the new time-integrated overlapped spectra and performs high-precision iterations of the phase polynomial fit on the filtered complex narrowband signals. The output of the PLL provides a new filtered and down-converted signal and its residual phase. The final residual phase in the stopped band is determined with respect to both sets of phase polynomials initially applied. The nature of the code allows filtering down the spacecraft tones as many times as desired in order to achieve milli-Hz accuracy. Depending on the signal-to-noise ratio of the detection, the output bandwidth of the phase detections after the PLL can range from a few kHz down to several milli-Hz. In the case of the VEX carrier line, this bandwidth is in the range of $10$--$100\,$Hz, which is sufficient for the IPS study and for the extraction of the residual phase of the VEX signal in a bandwidth of $20\,$Hz around the carrier line. The Doppler noise detection resolution ranges from one to one hundred mHz. 

Further description of the signal processing by means of a number of standard Matlab~\citep{matlab} procedures to estimate the IPS is described in Sect.~\ref{sec:res}.

\section{Results and analysis}\label{sec:res}
\begin{figure}[!ht]
 \begin{center}
 \includegraphics[width=250pt,clip]{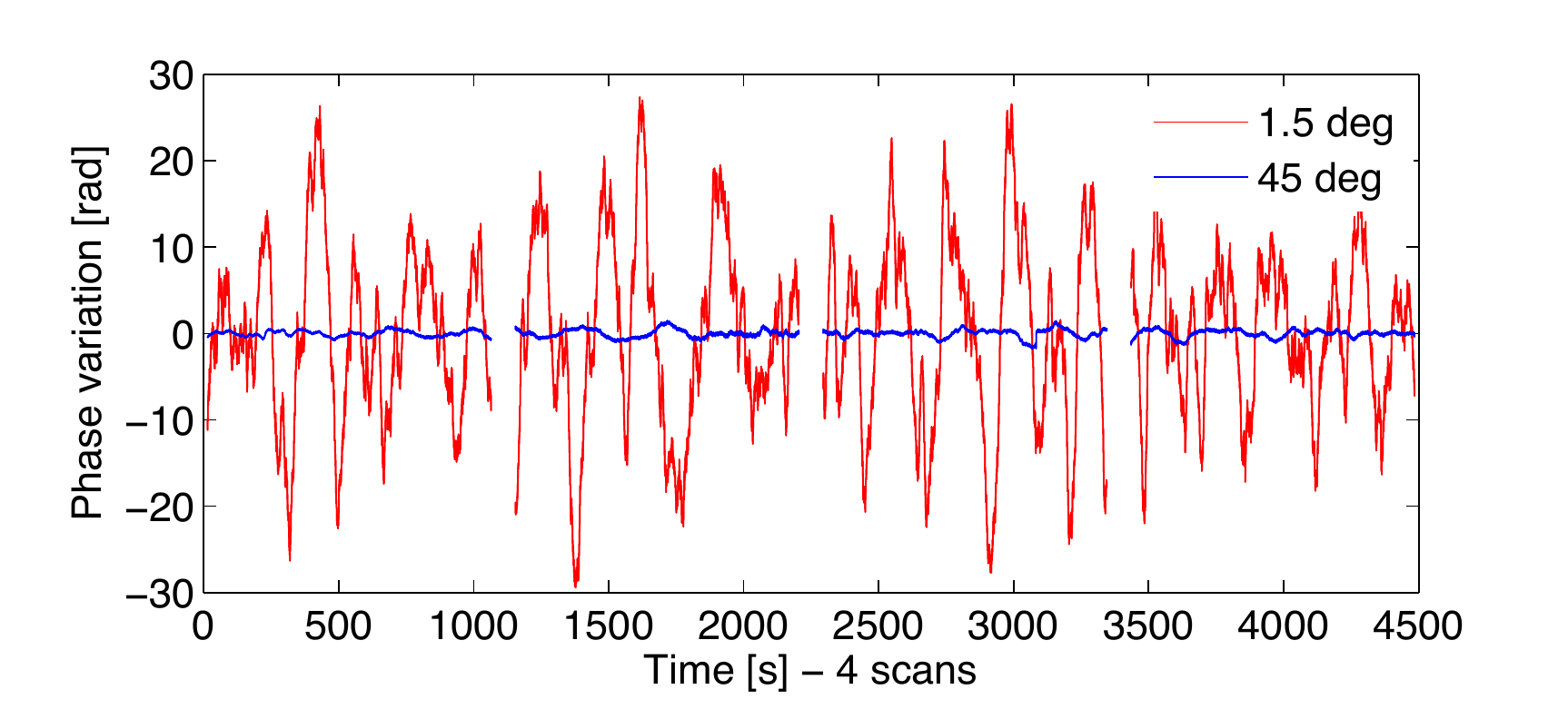}
 \end{center}
 \caption{Phase fluctuations of the VEX signal observed at Mets\"ahovi at low solar elongation ($1.5^\circ$, $2010$ Jan $16$, red line) and at Onsala radio telescope at wide angle ($45^\circ$, $2010$ Oct $30$, blue line). The system noise contribution in the scintillation band is negligible.}
 \label{fig:phases}
\end{figure}

We calculated the amount of variation present in the phase of each scan. The signal, with a stable phase, suffered fluctuations in the up- and downlinks due to interplanetary plasma. We noticed that these fluctuations directly depend on the solar elongation, on the distance from the Earth to the target, and on the activity of the solar wind. The phase fluctuation of the VEX spacecraft signal observed at low and high solar elongation epochs are compared in Fig.~\ref{fig:phases}. As expected, the level of phase fluctuation is $50$ times stronger at the lower solar elongation.

\begin{figure}[!ht]
 \begin{center}
  \includegraphics[width=250pt,clip]{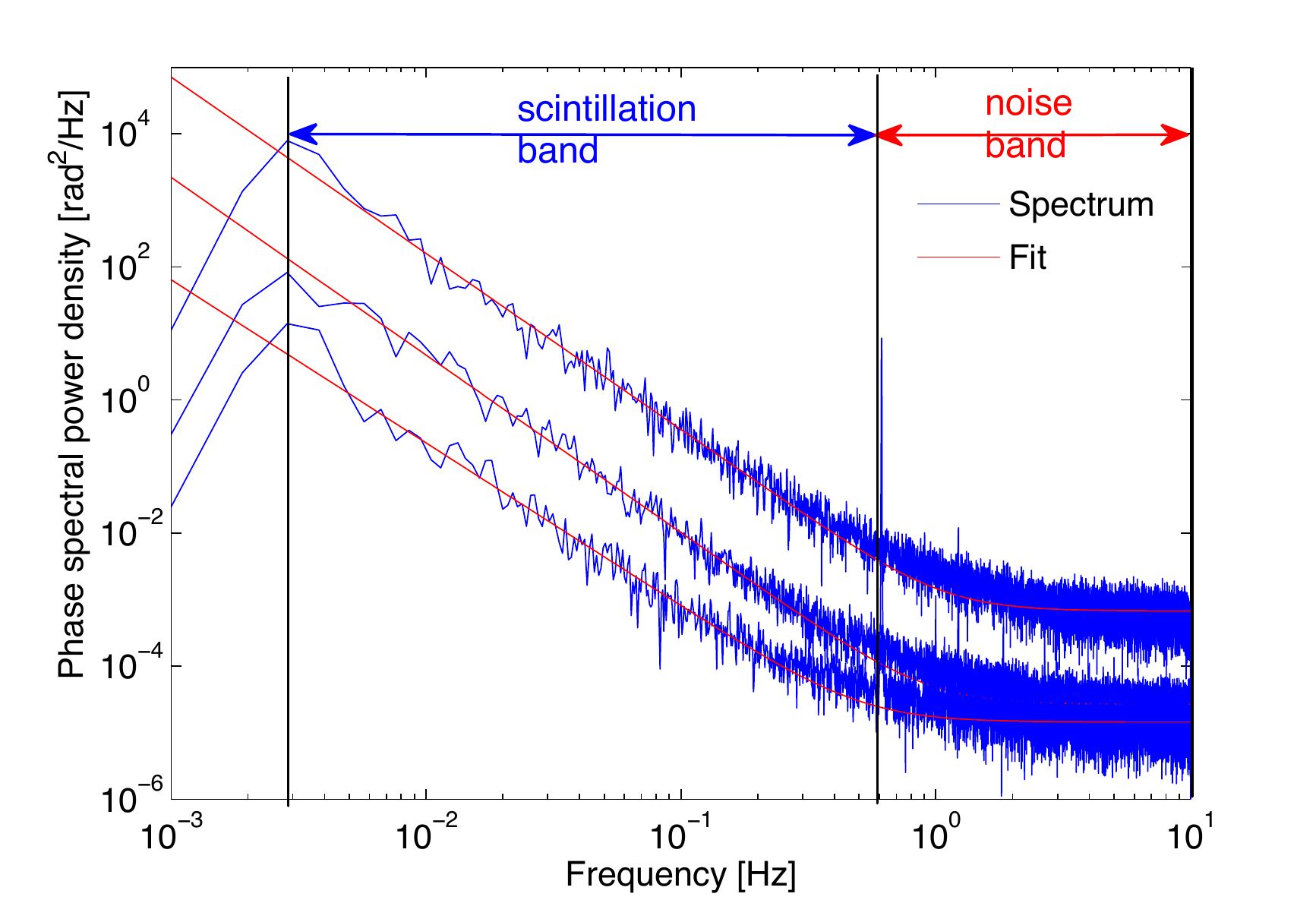}
 \end{center}
 \caption{Spectral power density of the phase fluctuations of VEX signal observed on $2010$ Jan $25$ ($3.4^\circ$, $1.72\,$AU, upper), $2010$ Mar $21$ ($19.4^\circ$, $1.58\,$AU, middle), and $2011$ Mar $21$ ($37.2^\circ$, $1.72\,$AU, lower) at Wettzell. Excessive system noise level at a close solar elongation is due to a leakage of Sun emission through the antenna sidelobes.}
 \label{fig:spectra}
\end{figure}

To characterise the nature of the phase variability and its intensity, we computed the spectral power density for each $19$-minute scan using the standard Matlab FFT libraries for discrete series. The spectral power densities for all scans were then averaged for the entire session. We determined the spectral index ($m$) in order to compare them to a Kolmogorov-type spectrum. We defined two boundaries for the scintillation band in our data as shown in Fig.~\ref{fig:spectra}. The lower limit ($f_\mathrm{min}$) is not constrained by the length of the scan ($1140\,$seconds or $1\,$mHz), but by the actual cut-off frequency determined by the order of the polynomial fit used for compensating for the bulk Doppler shift. In this case, $f_\mathrm{min}$ is at a level of $3\,$mHz, which represented the effective coherent integration time of $300\,$s. The upper limit ($f_\mathrm{max}$) depended on the amount of fluctuation and the noise of the receiver system. It is selected by visual inspection of the spectrum. The $f_\mathrm{max}$ usually ranges from $0.1$ to $0.5\,$Hz.

The fit in our model also includes the system noise. The system noise is calculated as the average of noise introduced in the range from $4$ to $7\,$Hz. That is correct in most cases, except at very low solar elongations. In that case, the data are dominated by IPS, and the noise is determined at higher frequencies (above $10\,$Hz).

To disentangle the scintillation and the noise contribution, a first-order approximation of the spectral power density on logarithmic scale ($L_\mathrm{ps}=a + m \cdot L_{f}$) was built. This allowed us to characterise the scintillation level caused by the interplanetary plasma. Here, $L_\mathrm{ps}$ is the average-windowed spectral power, $L_{f}$ the frequency on logarithmic scale, $m$ the slope, and $a$ is a constant. A fit was considered valid if its residual did not exceed $10\%$ of the total spectral power density.

The spectral power densities estimated at three different epochs with the same radio telescope are shown in Fig.~\ref{fig:spectra}. The observations were conducted with the Wettzell radio telescope when Venus was at solar elongations of $3.4$, $19.4$, and $37.2$ degrees, and at distances of $1.72$, $1.58$, and $1.19\,$AU from the Earth, respectively. The slopes of the spectral power densities were respectively $-2.388$, $-2.662$, and $-2.437$. These values agree with the spectral index of $-2.45$ found by~\citet{Woo}.

Typical values of the spectral index within the scintillation band  in our measurements ranged from $-2.80$ to $-2.12$. The average spectral index of these hundred observing sessions was $-2.42\,\pm\,0.25$. Following~\citet{Woo}, $m$ is defined as $m=1-p$, where $p$ is the Kolmogorov spectral index ($11/3$). Our values of $m \approx -8/3$ are consistent with a Kolmogorov power spectrum of fluctuations~\citep{Kolmogorov}. Such an approximation of the scintillation spectrum allowed us to characterise the scintillation contribution given by Eq.~\ref{eq:fluctuations}, as

\begin{equation}
 \sigma_\mathrm{Sc} = \left[-\frac{D(f_{0})\cdot f_{0}}{m+1}\right]^{\frac{m+1}{2}}.
 \label{eq:sigma_Sc}
\end{equation}

\begin{figure}[!ht]
 \begin{center}
  \includegraphics[width=250pt,clip]{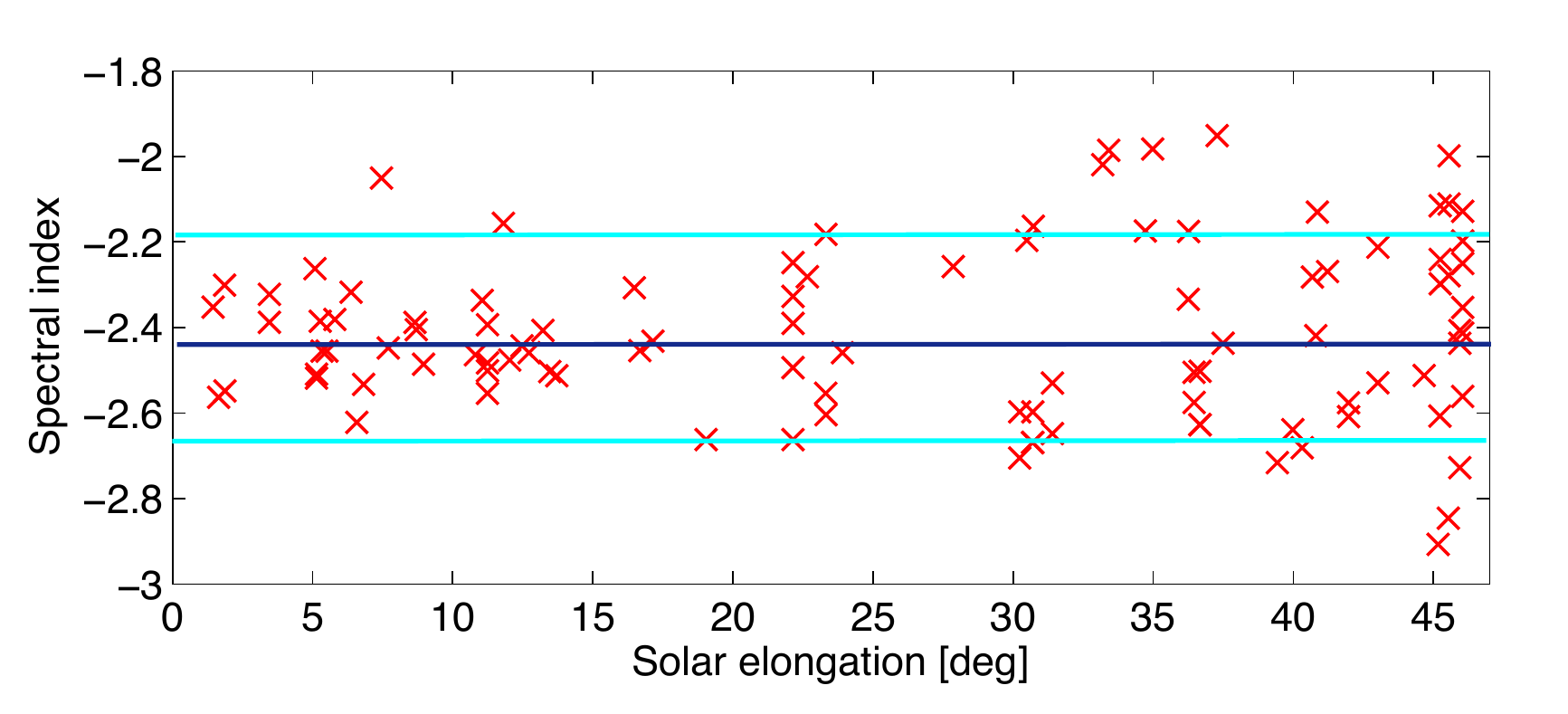}
 \end{center}
 \caption{Slopes of the spectral power density as a function of the solar elongation. Horizontal lines indicate the mean (in purple) and standard deviation (in cyan).}
 \label{fig:indices}
 \end{figure}

\noindent We studied the values of the spectral index to determine their dependence on a number of factors. We did not find any correlation between spectral index values and epochs, radio telescopes, solar elongations, or distances to the target. Our calculated indices are consistently similar during all the observations, regardless of the scintillation level. We conclude that the slope does not depend on these factors. As an example, Fig.~\ref{fig:indices} shows no dependence between solar elongation and the spectral index of the power density fluctuations.

We also considered the fact that the radio telescopes involved in the observations have different sensitivities and sizes. Increasing the number of participating antennas reduces the risk of obtaining biased results by a non-optimal performance of one of the telescopes. For instance, Mets\"ahovi is one of the stations with the highest SEFD and the most used radio telescope in this research. The spectrum of the phase scintillation observed at Mets\"ahovi showed a high variance when the spacecraft was at higher solar elongations. At these large angles, the phase fluctuations are low and generally dominated by the excessive system noise temperature of the receiver. However, antennas with better performance and higher sensitivity than Mets\"ahovi, such as Wettzell, Medicina, or Yebes, showed similar results. This confirmed that the Mets\"ahovi data were suitable for our study.

The consistency of the results obtained at each radio telescope can be verified in two different ways. The first relies on simultaneous observations with multiple radio telescopes. The other method is based on single antenna observations at consecutive epochs. The consistency of the results is illustrated in Table~\ref{tab:sessions}. Both situations were investigated thoroughly in the course of this study. Apart from a few inconsistent results that could be attributed to variations in weather conditions, no major incongruences were found.

\begin{table}[!ht]
 \caption{Results of two sessions carried out at the same epoch (first $2$ lines). Results of two consecutive days observed with the same radio telescope (last $2$ lines).}
 \begin{center}
 \begin{tabular}{l|c|c|c|c}
 \small
 Station		&	Date			&	Slope	&	$\sigma_\mathrm{Sc}$ [rad] 	&	$\sigma_\mathrm{n}$ [rad] \\
 \hline \hline
 Mets\"ahovi	&	$2011.03.27$	&	$-2.175$	&	$0.218$	&	$0.035$ 	\\
 Wettzell	&	$2011.03.27$	&	$-2.334$	&	$0.130$	&	$0.009$ 	\\
 Mets\"ahovi	&	$2010.05.17$	&	$-2.196$	&	$0.182$	&	$0.034$ 	\\
 Mets\"ahovi	&	$2010.05.18$	&	$-2.163$	&	$0.167$	&	$0.033$ 	\\
 \end{tabular}
 \end{center}
\label{tab:sessions}
\end{table}

The phase scintillation index ($\sigma_\mathrm{Sc}$) is the variance of the phase fluctuations within certain scintillation boundaries. The lower limit is again constrained by the cut-off frequency ($0.003$) determined by the order of the polynomial used for Doppler compensation. The upper limit is set arbitrarily to $3.003\,$Hz, for a total bandwidth of $3\,$Hz. It can be extrapolated from Eq.~\ref{eq:TEC} that the results depend weakly on the upper boundary. We therefore decided to always keep the upper boundary the same during all the observations, regardless of the amount of IPS.

The behaviour of the phase scintillation index was studied with respect to the solar elongation and the distance to the target. The solar elongation takes the closest point that the radio waves travel with respect to the Sun into consideration. The highest values of $\sigma_\mathrm{Sc}$ were obtained when Venus was at superior conjunction and solar elongation was close to $0^{\circ}$. For instance, Mets\"ahovi observed the spacecraft on $2010$ Jan $17$ when the spacecraft was at a solar elongation of $1.4$\,degrees and at a distance to the Earth of $1.71\,$AU. The measured $\sigma_\mathrm{Sc}$ was $40\,$rad. On the other hand, the measurements showed lower values when Venus was near inferior conjunction. The lowest $\sigma_{Sc}$ was achieved on $2011$ Mar $21$ at Wettzell with a value of $0.117\,$rad. The spacecraft was at $38\,$degrees of elongation and $1.19\,$AU of distance. 

\begin{figure}[!ht]
 \begin{center}
  \includegraphics[width=250pt,clip]{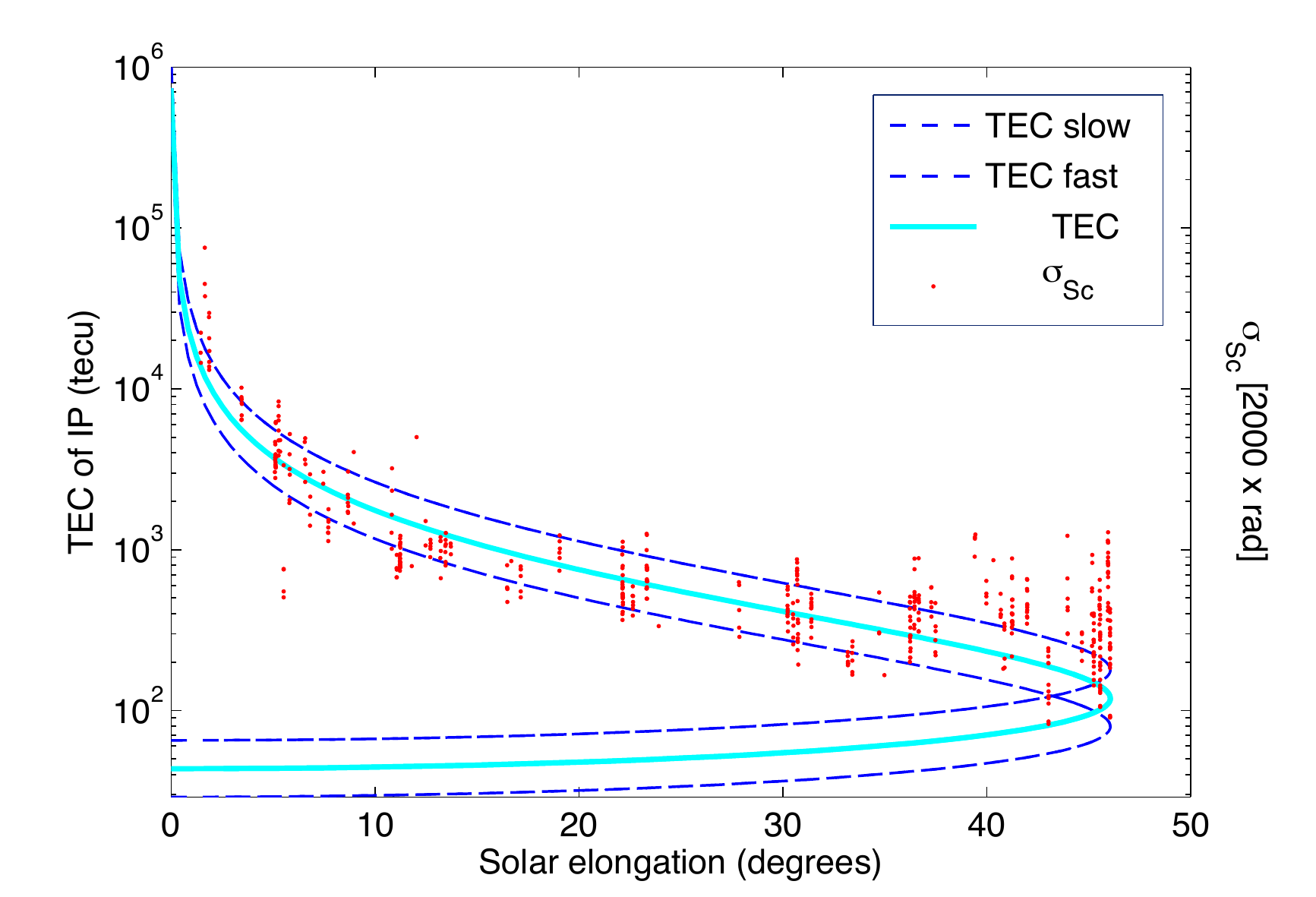}
 \end{center}
 \caption{Comparison of the phase scintillation indices onto Venus-to-Earth TEC along the line of sight. The $\sigma_{Sc}$ were multiplied by $2000$ to match TEC values (only 1-way link contribution shown). The red dots show the scaled $\sigma_{Sc}$ values. The TEC, taking into account fast winds (lower dark blue line, $3\,$electrons per cm$^{-3}$), and slow winds (upper dark blue line, $10\,$electrons per cm$^{-3}$), is plotted.}
 \label{fig:TECvsIPS}
\end{figure}

The phase scintillation indices as a function of the solar elongation were compared to the simulated TEC for the Venus orbit around the Sun. We used the approximation introduced in Eqs.~\ref{eq:elec_density} and~\ref{eq:TEC} to build the estimated TEC. To compare the scintillation with the TEC, the values of $\sigma_{Sc}$ have been scaled using a least squares minimisation method. The scaling coefficient is $0.00025$ radians/$tecu$. In other words, each $4000\,tecu$ of TEC contributes in $1\,$rad to $\sigma_{Sc}$ in the X-band with an integration time of $300\,$s. The goodness of the fit between the scaled phase scintillation indices, and the TEC was $0.28$ on logarithmic scale. To construct the TEC in Fig.~\ref{fig:TECvsIPS}, an average of five electrons per cm$^{-3}$ was used. TEC is also included as a reference in the cases for fast and slow winds, as was mentioned in Sect.~\ref{sec:the}. 

Based on Eq.~\ref{eq:sigma_Sc}, it is possible to characterise the expected phase variation of observations at different solar elongations by the following formula:

\begin{equation}
  \sigma_\mathrm{expected} = \frac{TEC}{4000} \cdot \left(\frac{8.4\,\mathrm{GHz}}{f_\mathrm{obs}}\right) \cdot \left(\frac{\tau}{300\,\mathrm{s}}\right)^\frac{m+1}{2}\,\,\mbox{[rad]},
 \label{eq:sigma}
\end{equation}

\noindent where TEC is computed using Eq.~\ref{eq:sigma_Sc}, $f_\mathrm{obs}$ is the observing frequency, and $\tau$ the nodding cycle for phase-referencing observations. It is worth noticing that one radian contribution per each $4000\,tecu$ in X-band corresponds to $1\,tecu$ fluctuative part independently of the solar elongation.

The contribution of the Earth's ionosphere is comparable to the interplanetary plasma's, when Venus is at inferior conjunction (as modelled in Fig.~\ref{fig:TECvsIPS}). However, the lack of observations conducted when Venus was close to Earth does not allow us to draw definitive conclusions. We calculated the estimated TEC of the ionosphere along the line of sight according to the procedure described in~\citet{Duev}. The TEC for the ionosphere was added to the TEC of the propagation in the solar system ($TEC = TEC_\mathrm{solar wind} + TEC_\mathrm{ionosphere}$). The improvement on the goodness of the fit between the phase scintillation indices and the TEC, taking the contribution of the ionosphere into account, is about $6.4\%$.

\section{Conclusions}\label{sec:con}
We estimated the spectral power density of the phase fluctuations at different solar elongations. They showed an average spectral index of $-2.42\,\pm\,0.25$, which agrees with the turbulent media described by Kolmogorov. From all our measurements, the slope of the phase fluctuations appears to be independent of the solar elongation.

Interplanetary scintillation indices calculated from the spacecraft phase variability were presented in this paper. The phase scintillation indices were measured for two Venus orbits around the Sun. The results obtained here provide a method for comparing the total electron content at any solar elongation and distance to the Earth with the measured phase scintillation index. The TEC values and phase scintillation present an error of $0.28$ on logarithmic scale.

Our measurements can be improved by considering the contribution of the Earth's ionosphere in our analysis. The improvement by including the ionosphere data is calculated to be of the order of $6.4\%$.

This study is also important for estimates of the spacecraft's state vectors using the VLBI phase-referencing technique. An important factor for VLBI phase-referencing is to select an optimal nodding cycle between the target and reference source. The results obtained from Venus Express offer precise information on the VLBI requirements and estimate the scintillation level at any epoch.

The results of this study are applicable to future space missions. It intends to be the basis for both calibration of state vectors of planetary spacecraft and further studies of interplanetary plasma with probe signals. 

\begin{acknowledgements}
This work was made possible by the observations conducted by a number of EVN radio telescopes and the geodetic station of Matera (Italy). The EVN is a joint facility of European, Chinese, Russian, and South African institutes funded by their national research councils. The authors would like to thank the station operators who conducted the observations at each of the stations; P. Boven and A. Kempeima (JIVE) for data transferring and processing. Also we would like to express gratitude to T. Morley (ESOC) and M. P\'erez-Ay\'ucar (ESA) for providing assistance for obtaining orbit and operations information on Venus Express. G.~Cim\`{o} acknowledges the EC FP7 project ESPaCE (grant agreement 263466). G. Molera~Calv\'es, and T.~M Bocanegra-Baham\'on acknowledges the NWO--ShAO agreement on collaboration in VLBI. W.~H. Wang acknowledges the NSFC project (U1231116).
\end{acknowledgements}


\end{document}